# Trade-off between interface morphology and compositional homogeneity in AlGaAs/GaAs quantum wells revealed by multislice electron ptychography

Yiwei Ju[1,†], Han-Hsuan Wu[2,†], Amberly Ricks[3], Levi Brown[2], Moaz Waqar[1], Rithvik Ramesh[3], Xingxu Yan[1], Toshihiro Aoki[4], Seth R. Bank[3], Xiaoqing Pan[1,2,4,*]

[1]Department of Materials Science and Engineering, University of California, Irvine, CA, USA

[2]Department of Physics and Astronomy, University of California, Irvine, CA, USA

[3]Department of Electrical and Computer Engineering, University of Texas Austin, Austin, Texas, USA

[4]Irvine Materials Research Institute, University of California Irvine, Irvine, CA, USA

† These authors contributed equally

*Corresponding author: xiaoqinp@uci.edu

## Abstract

AlGaAs/GaAs asymmetric coupled quantum wells (ACQWs) are promising platforms for enhanced second-order optical nonlinearities, with their performance strongly influenced by the structural quality of heterostructures. Flat interfaces, together with high compositional homogeneity, are generally desirable for optimizing device functionality. Here we show that sharp and flat AlGaAs/GaAs interface morphology dominates the second harmonic generation (SHG) response in AlGaAs/GaAs ACQWs, even when significant compositional fluctuations in Al/Ga occupancy are present within the AlGaAs layers. Using multislice electron ptychography, we resolve the three-dimensional (3D) interface morphology and compositional distribution in AlGaAs/GaAs ACQWs under distinct growth interruptions. Longer growth interruptions sharpen and flatten the AlGaAs/GaAs interfaces, but at the same time cause stronger compositional fluctuations in the AlGaAs layers. These findings clarify how growth interruption tunes quantum wells structures and provide direct guidance for quantum wells design in next-generation optoelectronic devices.

AlGaAs/GaAs quantum wells provide a versatile platform for optoelectronic and photonic devices, ranging from laser diodes [1-4], photodetectors [5-7], and nonlinear optical systems [8-11]. Enhanced interband nonlinearities have recently been predicted in AlGaAs/GaAs ACQWs when the quantum wells thickness, wells asymmetry, and barrier thickness are simultaneously optimized [11,12]. Several experimental studies have therefore sought to identify the ideal status by tuning the GaAs and AlGaAs layer thickness [13-16] and optimizing the AlGaAs composition [17-20].

Two structural parameters directly influence the optimization: the morphology of the AlGaAs/GaAs interfaces and the compositional homogeneity of the AlGaAs layers. Both parameters modify the local quantum wells asymmetry and thickness, thus influencing the second-order nonlinear susceptibility. Atomically sharp and flat interfaces are generally considered essential because interface roughness can broaden the local potential profile and spatially smear the nonlinear optical response [21-24]. Compositional uniformity in the AlGaAs layers is likewise expected to stabilize the band structure and suppress spatial variations in the nonlinear response [22]. As a result, epitaxial growth strategies often aim to minimize both interface roughness and compositional fluctuations [25-28]. However, the growth conditions that improve one structural parameter can degrade the other, making it challenge to optimize both parameters simultaneously.

Growth interruption has been widely utilized to improve interface morphology in heterostructures [29,30]. By allowing additional adatom diffusion at the growth front, it promotes sharper and flatter heterointerfaces. However, its impact on the compositional homogeneity, especially the uniformity of Al/Ga composition within the AlGaAs barrier, remains elusive. It is also unresolved whether interface morphology or compositional homogeneity dominates the nonlinear optical response in AlGaAs/GaAs ACQWs. Resolving these two structural contributions simultaneously is therefore necessary for understanding how growth interruption controls nonlinear functionality.

In this work, we reveal how AlGaAs/GaAs interface roughness and AlGaAs compositional homogeneity evolve with growth interruptions and determine their contributions to optical nonlinearity in AlGaAs/GaAs ACQWs. Using aberration-correction scanning transmission electron microscopy (STEM) and multislice electron ptychography (MEP), we resolve buried interface morphology and compositional variations in three dimensions. Comparison across three growth interruption conditions shows that longer growth interruption produces sharper AlGaAs/GaAs interfaces, but also stronger chemical inhomogeneity in the AlGaAs layers. Together with SHG measurements, these observations indicate that interface morphology is the dominant structural factor controlling optical nonlinearity in these quantum wells.

Three AlGaAs/GaAs ACQWs (16 periods) with different growth interruption times (0 s, 30 s, 90 s) were grown on the GaAs substrate (**Fig. 1a**) via molecular beam epitaxy (see **Supporting Information Note**) [29]. High-angle annular dark-field (HAADF) STEM images confirm that all three ACQWs retain high structural integrity across the entire heterostructure (**Fig. 1b** and **Fig. S1**). Individual ACQW periods, with thicknesses of approximately 10 nm, exhibit high crystalline quality and no observable defects (**Fig. 1c**). Energy-dispersive X-ray spectroscopy (EDS) further shows comparable Al, Ga, and As distributions across three growth interruption conditions (**Fig. 1d**). By maintaining similar crystalline quality, barrier thickness, and average composition, these samples provide a controlled platform for isolating how growth interruption modifies interface morphology and compositional homogeneity.

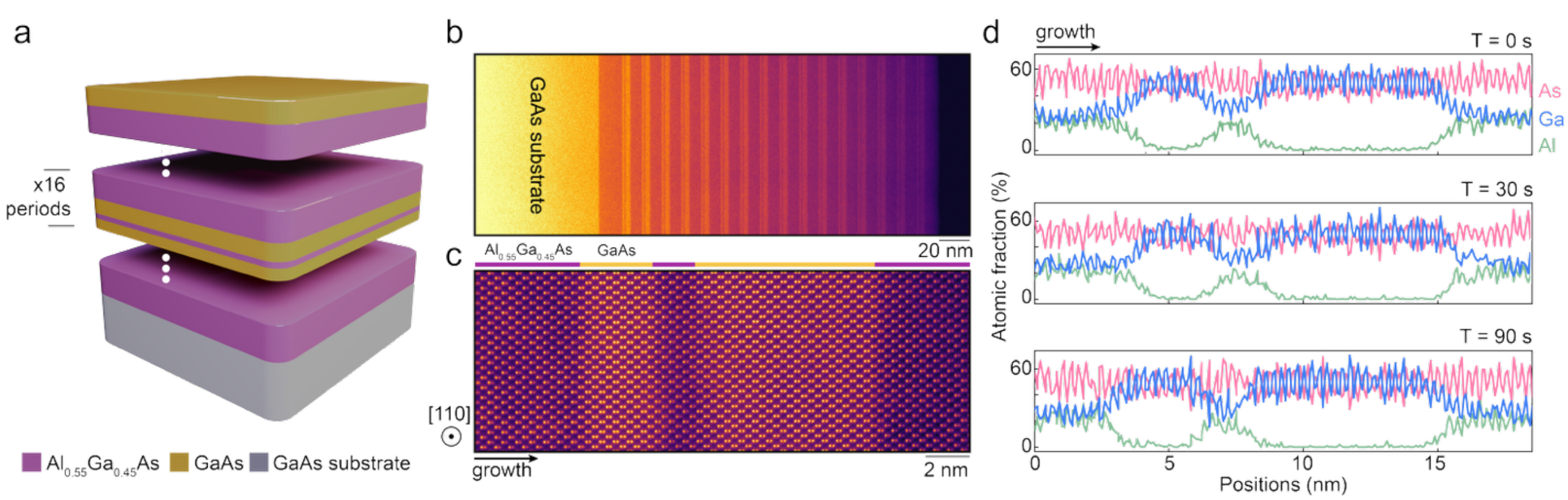

**Figure 1. Structure of AlGaAs/GaAs ACQWs.** (a) Schematic of the AlGaAs/GaAs ACQWs. The growth interruption at each interface is varied, while the other growth conditions are kept identical for three growth interrupted samples (see **Supporting Information Note** and **Fig. S2**). (b) Large field-of-view HAADF-STEM image showing the layered heterostructure and uniform quantum well periodicity. (c) Zoomed-in HAADF-STEM image of a representative ACQW period,

confirming high crystalline quality. (d) EDS line profiles across one ACQW period, where ($T$) denotes the growth interruption time. The Al, Ga, and As distributions show no significant variation among three growth interrupted samples, confirming consistent average composition across the series (see raw datasets in **Fig. S3**).

To uncover 3D interface morphology and compositional homogeneity, we employ MEP, which enables phase-sensitive reconstruction with sub-angstrom resolution [31,32]. In conventional HAADF-STEM images, the contrast scales approximately with $Z^{1.7}$ [33], where $Z$ denotes the atomic number, and represents a projection of the scattering potential along the beam direction. As a result, depth-dependent structural variations are averaged into a two-dimensional (2D) image and are difficult to resolve. In contrast, MEP reconstructs the phase of the electron wave, providing access to depth-dependent variations in the sample potential. The experimental configuration is illustrated in **Fig. 2a**, where a coherent electron probe is scanned across the sample and a diffraction pattern is recorded at each scan position, generating a four-dimensional (4D) dataset. From this dataset, both amplitude and phase information of samples can be reconstructed (**Figs. S4-7**). One representative reconstructed MEP phase image of 30s-interruption is displayed in **Fig. 2b**, providing enhanced structural detail compared to conventional HAADF-STEM imaging (**Figs. S8-9**). More importantly, the MEP resolves variations along the beam direction. As shown in **Fig. 2c**, phase profiles along the depth can be extracted. In the region I, the phase of Al/Ga sites remains that same along the depth direction since it is the pure GaAs layer. However, in regions II to IV, the phase intensity of Al/Ga atomic column is weaker due to the reduced average atomic number and exhibit obvious variation along depth direction and depending on the atom position. The atomic columns across the AlGaAs/GaAs interface show significant variations along the depth. The interface is therefore not an atomically abrupt boundary; instead, it exhibits a bumpy, spatially nonuniform morphology.

With the capability of resolving variations along depth direction, we then extract the phase of Al/Ga atomic columns across all reconstructed slices (**Fig. 2d and Fig. S11**). The normalized phase of Al/Ga atomic columns, $\varphi(\boldsymbol{r})$, maps the 3D spatial distribution of AlGaAs and GaAs regions (**Fig. 2e and Fig. S12**). Since the phase contrast is roughly proportionally to $Z$, we may use the relative changes in the phase contrast to measure elemental variation at every atom position. The following two parameters are further measured to determine the interface morphology and compositional homogeneity across the growth interruption series. i) The gradient of the normalized phase at Al/Ga atomic columns, $|\nabla\varphi(\boldsymbol{r})|$, identifies the interface locations and width (**Fig. 2f and Fig. S13**). ii) The local variance of the normalized phase at Al/Ga atomic columns, $\sigma_{\varphi}(\boldsymbol{r})$, captures local fluctuation, especially within the AlGaAs layer (**Fig. 2g and Fig. S14**), serving as a measure of compositional homogeneity.

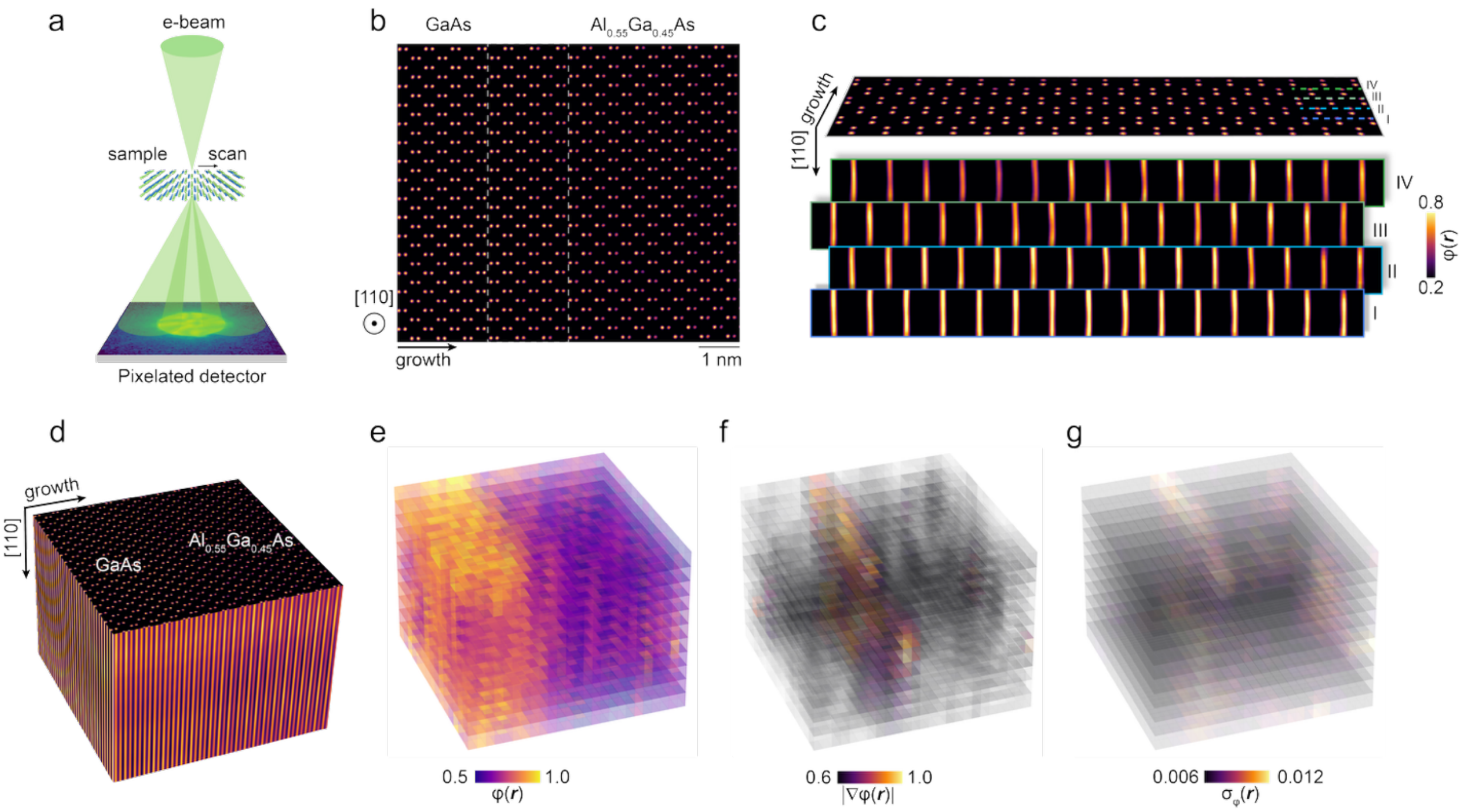

**Figure 2. MEP resolves 3D structure of AlGaAs/GaAs ACQWs.** (a) Schematic of the MEP experiment, in which a coherent electron probe is scanned across the sample and a diffraction pattern are recorded at each probe position. (b) Reconstructed ptychographic phase image of the AlGaAs/GaAs ACQWs. (c) Upper panel: zoomed-in image of interfacial region marked by the white dashed box in (b). Lower panel: depth-dependent phase profiles of four adjacent Al/Ga atomic columns across the interface. The inhomogeneity along the depth indicates a spatially nonuniform, bumpy interface morphology. (d) 3D phase image of the AlGaAs/GaAs ACQWs reconstructed from multislice electron ptychography. (e) 3D spatial map of normalized phase at Al/Ga atomic columns, $\varphi(\boldsymbol{r})$, which tracks the spatial distribution of the AlGaAs and GaAs regions. (f) 3D spatial map of the gradient of normalized phase at Al/Ga atomic columns, $|\nabla\varphi(\boldsymbol{r})|$, used to identify the interface position and width. (g) 3D spatial map of local variance of normalized phase at Al/Ga atomic columns, $\sigma_{\varphi}(\boldsymbol{r})$, used to determine the compositional homogeneity.

A step of growth interruption, which the chamber was evacuated before switching to the consecutive precursor gas, was intentionally added to sharpen and flatten the AlGaAs/GaAs interface. **Figs. 3 (a–c)** present MEP phase images from three growth interrupted samples, each containing AlGaAs/GaAs interfaces. All samples consistently give rise to bumpy morphology (**Fig. S15**), but the spatial extent of the interfacial region changes with growth interruptions. 3D spatial map of gradient, $|\nabla\varphi(\boldsymbol{r})|$, in all slices identify the interfacial regions, across which the normalized phase changes most strongly. Compared with the 0 s growth interrupted sample, the 30 s sample shows a narrower interfacial region (**Figs. 3d-e**). Histograms of the normalized phase at Al/Ga atomic columns from all slices confirming this narrowing. By using S-shape function to fit the histogram profile, it yields interfacial widths of approximately 4.3 nm and 1.9 nm for the 0 s and 30 s samples (**Figs. 3g-h**). Additionally, growth interruptions make the interfacial morphology flat. The large $|\nabla\varphi(\boldsymbol{r})|$ is more tightly confined within 1~2 unit cells in the 30 s sample, whereas the 0 s sample retains a pronounced bumpy feature toward the AlGaAs side. Besides, the 90 s sample exhibits an interfacial width of approximately 1.8 nm (**Fig. 3f and i**), which is very similar to that of the 30 s sample, and both interfaces are considerably sharper than that of the 0 s sample. However, locally enhanced $|\nabla\varphi(\boldsymbol{r})|$ appears in the AlGaAs layers of the 90 s sample, suggesting stronger local compositional fluctuations, as discussed in **Fig. 4**.

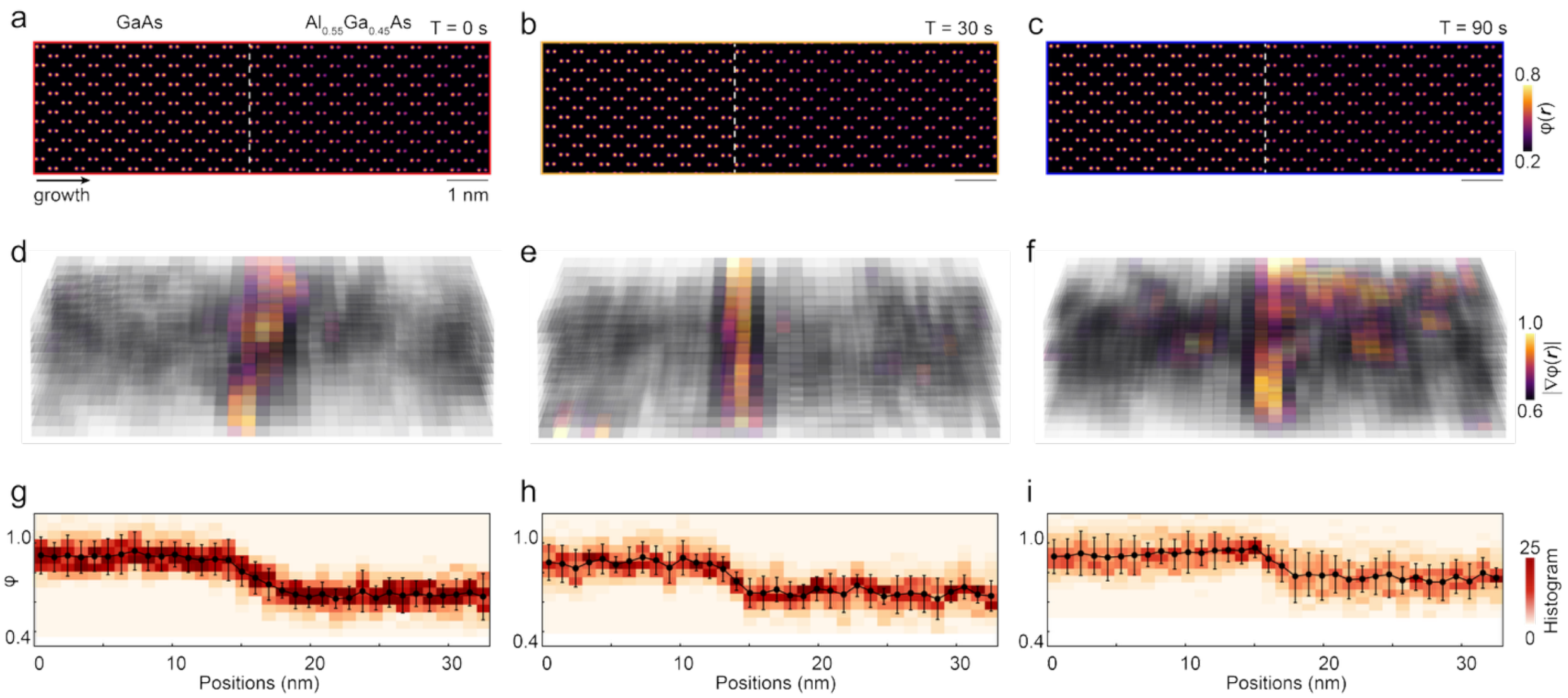

**Figure 3. Effects of growth interruptions on the interfacial morphology.** (a–c) Ptychographic reconstructed phase images of AlGaAs/GaAs ACQWs with growth interruption times of 0 s (a), 30 s (b), and 90 s (c). The white dashed lines indicate the approximate positions of the interfaces. (d–f) 3D spatial map of the gradient of normalized phase at Al/Ga atomic columns, $|\nabla \varphi(\boldsymbol{r})|$, extracted from regions shown in (a-c). The gradient signal highlights the interfacial regions and becomes more confined with growth interruption. (g–i) 2D histogram of the normalized phase, $\varphi(\boldsymbol{r})$, collected from all reconstructed slice for the 0s (g), 30s (h) and 90 s (i) samples (see raw datasets in **Fig. S12**). Circles mark the histogram expectation values, and error bars indicate the full width at half maximum. The narrowing of the distribution reflects the reduced interfacial width under growth interruption.

Longer growth interruptions also give rise to stronger compositional inhomogeneity within the AlGaAs layer. **Figs. 4 (a–c)** present the 3D spatial maps of normalized phase at Al/Ga atomic columns in AlGaAs layers for three growth interruption conditions. The phase of As columns was used to normalize the phase information slice by slice to allow a rational comparison across samples. Without growth interruptions (0 s), the phase of Al/Ga columns remains relatively uniform, indicating minimal compositional fluctuation. With increasing growth interruptions, the spatial variation of the phase of Al/Ga columns becomes progressively pronounced, with the 90 s sample showing clear inhomogeneity throughout the AlGaAs layers. Histograms collected from all slices confirms the largest full width at half-maximum (FWHM) in the 90 s samples (**Figs. 4g-i**), demonstrating the strongest compositional inhomogeneity in the 90 s sample.

Spatial maps of local variance, $\sigma_{\varphi}(\boldsymbol{r})$, further confirm the stronger compositional inhomogeneity with growth interruptions (**Figs. 4d-f**). $\sigma_{\varphi}(\boldsymbol{r})$ becomes stronger and more inhomogeneous as the growth interruption increases. Regions with large $\sigma_{\varphi}(\boldsymbol{r})$ appear at different lateral positions and depths, including interior slices as well as top and bottom slices. This distribution confirms that the compositional inhomogeneity is intrinsic rather than arising from TEM sample preparation, which would be expected to contribute more to stronger features to the top and bottom regions.

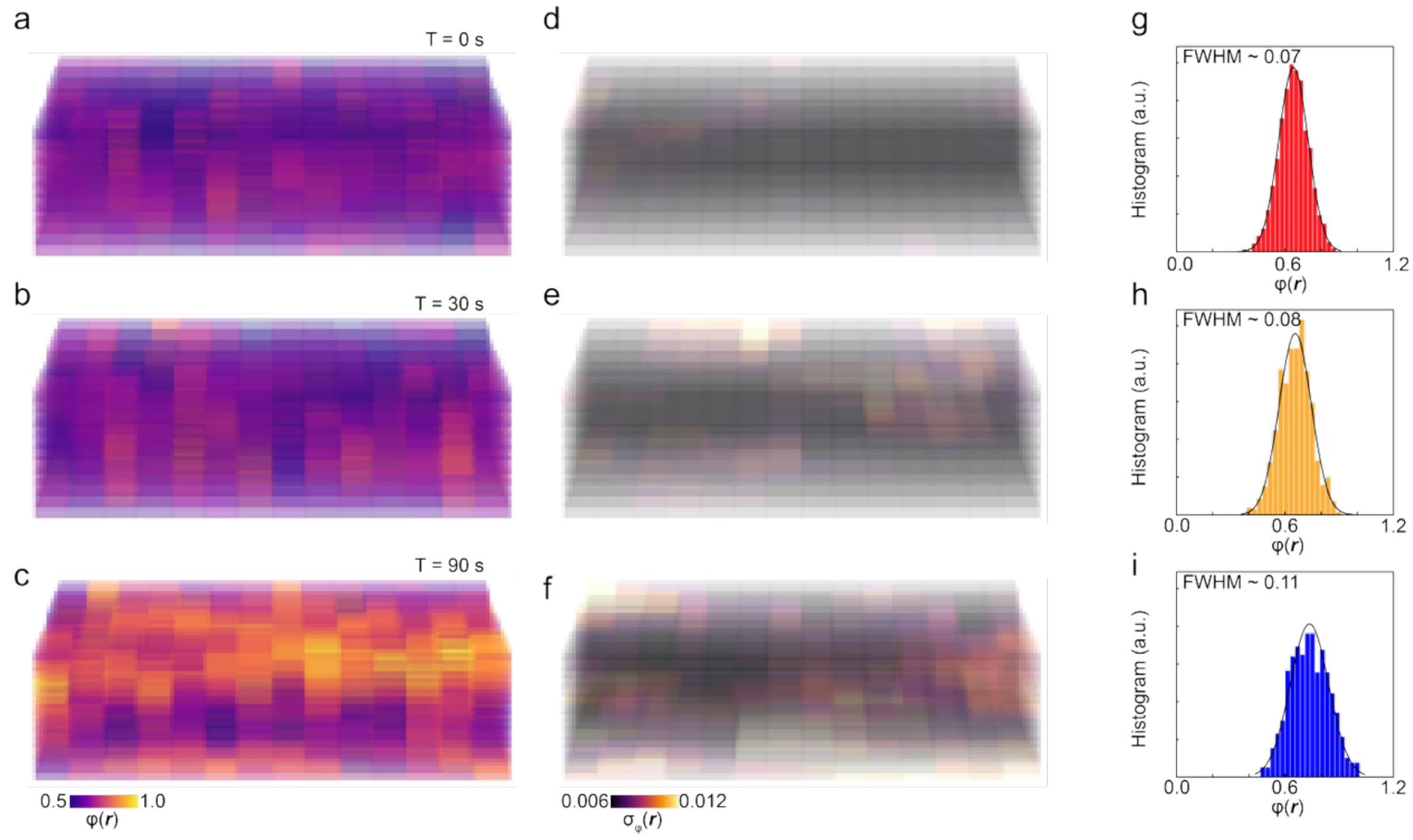

**Figure 4. Effects of growth interruptions on the compositional inhomogeneity within the AlGaAs layers.** (a–c) 3D spatial map of normalized phase at Al/Ga atomic columns, $\varphi(\boldsymbol{r})$, within the AlGaAs layers for three growth interrupted times of 0 s (a), 30 s (b), and 90 s (c). The phase is normalized to the As site in each slice for fair comparison across samples. (d–f) 3D spatial map of local variance of the normalized phase at Al/Ga atomic columns, $\sigma_{\varphi}(\boldsymbol{r})$, derived from data shown in (a–c). (g–i) Histogram of $\varphi(\boldsymbol{r})$ for Al/Ga sites in three growth interrupted samples. The FWHM is measured.

MEP thus reveals a structural trade-off induced by growth interruption: sharper and flatter AlGaAs/GaAs interfaces form at the expense of compositional homogeneity within the AlGaAs layers (**Fig. 5a**). The growth interruption assisted interface smoothing morphology is consistent with previous studies showing that growth interruption can promote surface relaxation and reduce interface roughness in AlGaAs/GaAs quantum wells [34], which is commonly attributed to enhanced adatom migration during the interruption. Without growth interruption, limited adatom mobility kinetically freezes the interface shortly after deposition, preserving a broader and bumpy morphology. Growth interruption allows additional lateral diffusion along the growth surface, leading to flatter interfaces and reduced interfacial width. At the same time, the enhanced mobility may facilitate local cation redistribution or intermixing within the AlGaAs layers, producing stronger compositional fluctuations. Therefore, interface morphology and compositional homogeneity evolve in opposite directions with growth interruption (**Fig. 5b**), making them challenging to optimize simultaneously through growth interruption.

The optical nonlinearity of these AlGaAs/GaAs ACQWs, however, is dominated by interface morphology. **Figure 5c** shows the measured SHG signals from three growth interrupted samples. The SHG signal increases with longer growth interruption. This trend indicates that the improvement in interface morphology outweighs the detrimental influence of compositional fluctuations, establishing interface roughness as the primary structural parameter controlling the nonlinear optical response in AlGaAs/GaAs ACQWs.

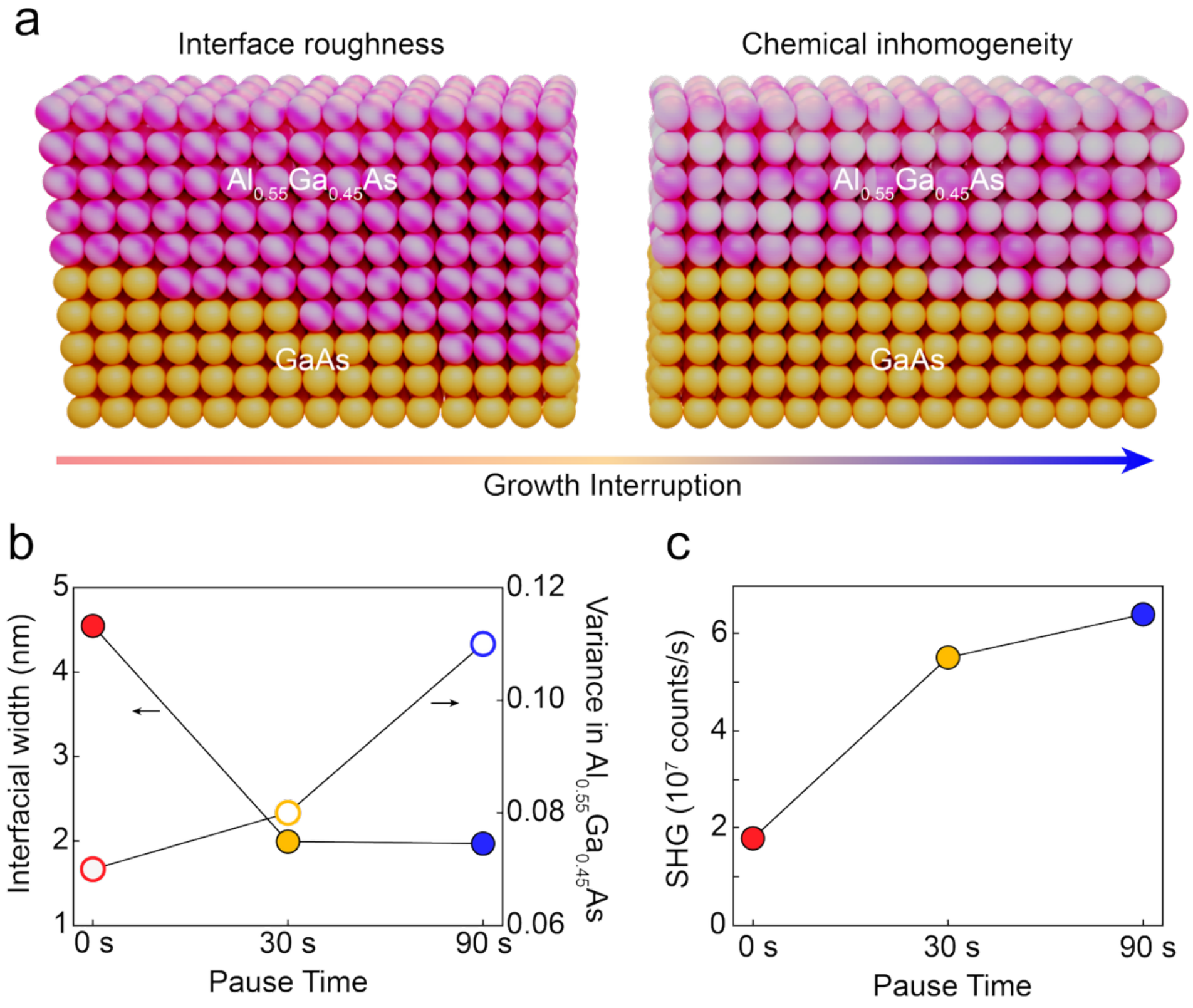

**Figure 5. Opposite variations of interfacial morphology and compositional homogeneity under growth interruptions and their contributions to the SHG response.** (a) Schematic illustration of the growth interruption effect on AlGaAs/GaAs ACQWs. Without growth

interruption, limited adatom mobility kinetically freezes a broader and bumpy interface while maintain relatively weak compositional inhomogeneity in the AlGaAs layers. With finite growth interruption, the enhanced adatom mobility promotes interface smoothing but also increases compositional inhomogeneity within the AlGaAs layers. (b) Evolution of interfacial width (left axis) and variance of normalized Al/Ga phase within the AlGaAs layers (right axis) as a function of growth interruptions. (c) SHG signals measured from three growth interrupted samples, showing that the enhanced nonlinear optical response follows the improvement in interface morphology. The data is summarized from ref. [29].

Interface morphology and compositional homogeneity are two key structural parameters for optimizing the optoelectronic functionality of AlGaAs/GaAs ACQWs. Resolving the dependence of interface morphology and compositional homogeneity upon growth strategies is difficult with conventional STEM imaging because relevant structural changes are usually subtle and easily overshadowed by projection-averaged signal in conventional STEM imaging. Depth-resolved imaging combined with high lateral spatial resolution is therefore essential for accessing the buried interface. Using MEP, we can resolve both the buried interface and subtle compositional fluctuations in AlGaAs/GaAs ACQWs, revealing opposite dependence of these two structural parameters upon growth interruptions.

Given the opposite requirement on the adatom mobility to smooth interface and minimize compositional fluctuation, achieving both a sharp interface and a homogeneous AlGaAs composition remains challenging [34-36]. Our results demonstrate how this trade-off affects optical nonlinearity in AlGaAs/GaAs ACQWs. Even when growth interruption produces measurable compositional inhomogeneity, the sharper and flatter interface morphology dominates the enhanced nonlinear optical response. These findings provide insightful guidance for quantum wells design for optoelectronic devices and provide a route for connecting buried 3D structure to optical functionality in quantum wells heterostructures.

**Acknowledgements**

The experimental work was supported by a Multidisciplinary University Research Initiative from the Air Force Office of Scientific Research (AFOSR MURI Award No. FA9550-22-1-0307). The authors acknowledge the use of facilities and instrumentation at the UC Irvine Materials Research Institute (IMRI), which is supported in part by the National Science Foundation through the UC Irvine Materials Research Science and Engineering Center (DMR-2011967). The authors thank Dr. Haozhi Sha for valuable discussion on electron ptychography.

**References**

[1] H. Zhu, K. Liu, C. Xiong, et al., "The effect of external stress on the properties of AlGaAs/GaAs single quantum well laser diodes," *Microelectronics Reliability* 55 (2015): 62-65.

[2] Y. Nagai, K. Shigihara, S. Karakida, et al., "Characteristics of laser diodes with a partially intermixed GaAs-AlGaAs quantum well," *IEEE journal of quantum electronics* 31 (2002): 1364-1370.

[3] A. A. Marmalyuk, M. A. Ladugin, A. Yu. Andreev, et al., "AlGaAs/GaAs laser diode bars with improved thermal stability," *Quantum Electronics* 43 (2013): 895-897.

[4] M. A. Ladugin, A. A. Marmalyuk, A. A. Padalitsa, et al., "Laser diode bars based on AlGaAs/GaAs quantum-well heterostructures with an efficiency up to 70%," *Quantum Electronics* 47 (2017): 291-293.

[5] X. Zhu, F. Lin, Z. Zhang, et al., "Enhancing performance of a GaAs/AlGaAs/GaAs nanowire photodetector based on the two-dimensional electron–hole tube structure," *Nano Letters* 20 (2020): 2654-2659.

[6] J. Lu, R. Surridge, G. Pakulski, et al., "Studies of high-speed metal-semiconductor-metal photodetector with a GaAs/AlGaAs/GaAs heterostructure," *IEEE transactions on electron devices* 40 (2002): 1087-1092.

[7] M. Bazalevsky, S. Didenko, S. Yurchuk, et al., "AlGaAs/GaAs photodetectors for detection of luminescent light from scintillators," *Journal of Physics: Conference Series*. 586 (2015): 012018.

[8] X. H. Qu, H. Ruda, S. Janz, et al., "Enhancement of second harmonic generation at 1.06 μm using a quasi-phase-matched AlGaAs/GaAs asymmetric quantum well structure," *Applied Physics Letters* 65 (1994): 3176-3178.

[9] P. Boucaud, F. H. Julien, D. D. Yang, et al., "Saturation of second-harmonic generation in GaAs–AlGaAs asymmetric quantum wells," *Optics Letters* 16 (1991): 199-201.

[10] C. Y. Cheng, "Second-harmonic generation by reflection from AlGaAs/GaAs structures, " The University of Texas at Arlington (1994).

[11] J. Khurgin, "Second-order nonlinear effects in asymmetric quantum-well structures," *Physical Review B* 38 (1988): 4056.

[12] R. Ramesh, T. Hsieh, A. M. Skipper, et al., "Interband second-order nonlinear optical susceptibility of asymmetric coupled quantum wells," *Applied Physics Letters* 123 (2023): 251111.

[13] P. Roussignol, M. Gurioli, L. Carraresi, et al., "Electron and hole tunneling times in GaAs/AlGaAs asymmetric double quantum well heterostructures," *Superlattices and Microstructures* 9 (1991): 151-155.

[14] E. M. Lopes, J. L. Duarte, L. C. Pocas, et al., "Exciton behavior in GaAs/AlGaAs coupled double quantum wells with interface disorder," *Journal of Luminescence* 130 (2010): 460-465.

[15] A. Torabi, K. F. Brennan, C. J. Summers, "Photoluminescence studies of coupled quantum well structures in the AlGaAs/GaAs system," *Quantum Well and Superlattice Physics* 792 (1987): 152 -156.

[16] J. M. Roberts 1997, Electronic properties of quantum wells for field effect transistor applications. University of London, University College London (United Kingdom).

[17] W. Shichi, T. Ito, M. Ichida, et al., "Dependence of electron g-factor on barrier aluminum content in GaAs/AlGaAs quantum wells," *Japanese Journal of Applied Physics* 48 (2009): 063002.

[18] I. A. Yugova, A. Greilich, D. R. Yakovlev, et al., "Universal behavior of the electron g factor in Ga As/$Al_xGa_{1-x}As$ quantum wells," *Physical Review B* 75 (2007): 245302.

[19] C. R. Hall 2011, Ultrafast dynamics in semiconductor quantum wells, Diss. Swinburne.

[20] P. Nithiananthi, G. Vignesh, "Theoretical study of exciton Mott transition in quantum wells through interband transition energy of excitons," *AIP Conference Proceedings* 2995 (2024): 020197.

[21] B. Yang, Y. Cheng, Z. Wang, et al., "Interface roughness scattering in GaAs–AlGaAs modulation-doped heterostructures," *Applied Physics Letters* 65 (1994): 3329-3331.

[22] T. Wang, H. Song, K. He, "Structural design and molecular beam epitaxy growth of GaAs and InAs heterostructures for high mobility two-dimensional electron gas," *Quantum Frontiers* 3 (2024): 13.

[23] H. Sakaki, T. Noda, K. Hirakawa, et al., "Interface roughness scattering in GaAs/AlAs quantum wells," *Applied Physics Letters* 51 (1987): 1934-1936.

[24] W. Braun, K. H. Ploog, "In situ technique for measuring Ga segregation and interface roughness at GaAs/AlGaAs interfaces," *Journal of Applied Physics* 75 (1994): 1993-2001.

[25] Y. Zhang, Y. Gu, P. Chen, et al., "Composition uniformity characterization and improvement of AlGaAs/GaAs grown by molecular beam epitaxy," *Materials Science in Semiconductor Processing* 79 (2018): 107-112.

[26] M. Ilegems, "Beryllium doping and diffusion in molecular-beam epitaxy of GaAs and $Al_x Ga_{1-x}As$," *Journal of Applied Physics* 48 (1977): 1278-1287.

[27] N. Chand, S. N. G. Chu, "Origin and improvement of interface roughness in AlGaAs/GaAs heterostructures grown by molecular beam epitaxy," *Applied Physics Letters* 57 (1990): 1796-1798.

[28] M. A. Herman, D. Bimberg, J. Christen, "Heterointerfaces in quantum wells and epitaxial growth processes: Evaluation by luminescence techniques," *Journal of Applied Physics* 70 (1991): R1-R52.

[29] R. Ramesh, M. Brown, A. Ricks, et al., "Enhanced Interband Optical Nonlinearities from Coupled Quantum Wells," *arXiv preprint arXiv*:2602 (2026).

[30] K. Kanamoto, K. Fujiwara, Y. Tokuda, et al., "Surface diffusion during MBE growth of GaAs-AlGaAs single quantum wells on vicinal surfaces," *Journal of Crystal Growth* 95 (1989): 273-276.

[31] Y. Jiang, Z. Chen, Y. Han, et al., "Electron ptychography of 2D materials to deep sub-ångström resolution," *Nature* 559 (2018): 343-349.

[32] Z. Chen, Y. Jiang, Y. T. Shao, et al., "Electron ptychography achieves atomic-resolution limits set by lattice vibrations," *Science* 372 (2021): 826-831.

[33] F. Li, M. J. Cabral, B. Xu, et al., “Giant piezoelectricity of Sm-doped Pb $(Mg_{1/3}Nb_{2/3})O_3$–$PbTiO_3$ single crystals,” *Science* 364, 264–268 (2019).

[34] A. Madhukar, T. C. Lee, M. Y. Yen, et al., "Role of surface kinetics and interrupted growth during molecular beam epitaxial growth of normal and inverted GaAs/AlGaAs (100) interfaces: A reflection high-energy electron diffraction intensity dynamics study," *Applied Physics Letters* 46 (1985): 1148-1150.

[35] Y. Horikoshi, M. Kawashima, H. Yamaguchi, "Migration-enhanced epitaxy of GaAs and AlGaAs," *Japanese Journal of Applied Physics* 27 (1988): 169.

[36] N. Chand, S. N. G. Chu, "Origin and improvement of interface roughness in AlGaAs/GaAs heterostructures grown by molecular beam epitaxy," *Applied Physics Letters* 57 (1990): 1796-1798.